\documentclass[mnsc,noblindrev]{informs3}

\RequirePackage{tgtermes}
\RequirePackage{newtxtext}
\RequirePackage{newtxmath}
\RequirePackage{bm}
\RequirePackage{endnotes}

\OneAndAHalfSpacedXII 

\usepackage{algorithm}
\usepackage{algpseudocode}
\usepackage{tikz}

\usepackage{natbib}
 \bibpunct[, ]{(}{)}{,}{a}{}{,}%
 \def\bibfont{\small}%

\usepackage{caption}
\usepackage[labelfont=sf]{subcaption}
\EquationsNumberedThrough    

\TheoremsNumberedThrough     
\ECRepeatTheorems  %

\MANUSCRIPTNO{}

\begin{document}


\RUNAUTHOR{Zhang, Zheng and Yan}

\RUNTITLE{Generative AI and Two-Tiered Online Mental Health Communities}

\TITLE{Generative AI and Two-Tiered Online Mental Health Communities}

\ARTICLEAUTHORS{%
\AUTHOR{Manyang Zhang}
\AFF{School of Management,
Beijing Institute of Technology, \EMAIL{manyangz@bit.edu.cn}}

\AUTHOR{Jinyang Zheng}
\AFF{Simon Business School,
University of Rochester, \EMAIL{zhengjy@rochester.edu}}

\AUTHOR{Zhijun Yan}
\AFF{School of Management,
Beijing Institute of Technology, \EMAIL{yanzhijun@bit.edu.cn}}
} 

\ABSTRACT{%
Online mental health communities (OMHCs) are tiered platforms that connect patients with licensed counselors through public Q\&A forums and paid private consultations. Their two-tier structure creates a strategic dilemma for genAI integration. On the one hand, conversational agents can provide scalable and timely responses to a broader set of patients, alleviating persistent supply shortages. On the other hand, their large-scale presence may reshape counselors’ participation in delivering nuanced expertise, emotionally sensitive support, and paid consultations—the ultimate sources of platform revenue and long-run sustainability.
Leveraging a quasi-natural experiment from the integration of a genAI-based conversational agent in a leading OMHC, we examine how AI entry affects counselor participation. Using multiple identification strategies, we find that posting intensity increases significantly after AI integration, while average response length remains unchanged and per-post social recognition declines. Mechanism analyses show that AI improves responsiveness and expands patient engagement, enlarging counselors’ opportunity sets, with activity partially reallocated from a nearby non-AI subforum. Counselors respond heterogeneously: intrinsically motivated counselors reduce participation, whereas economically motivated counselors intensify competitive effort. These dynamics generate cross-tier spillovers: inactive counselors experience declines in paid consultations, while those who increase public participation preserve or expand downstream demand.
Together, our findings show that in tiered professional platforms, demand expansion and competitive incentives can outweigh intrinsic crowding-out. Rather than simply substituting for human expertise, internal genAI reshapes participation by expanding engagement and activating strategic effort, allowing platforms to benefit simultaneously from automated scalability and sustained professional participation.
}%





\maketitle

\section{Introduction}\label{intro} 
Online mental health communities (OMHCs) are specialized two-sided platforms that connect patients seeking psychological support with licensed counselors through public Q\&A forums and paid private consultations. These platforms provide scalable and accessible alternatives to traditional clinical settings and play an increasingly important role in addressing widespread psychological needs \citep{RN101}. Yet OMHCs face a structural tension. While patient demand is highly responsive to timeliness and interaction quality, the supply of qualified counselors remains limited \citep{RN871}. Delayed responses and insufficient engagement can dampen participation, hinder trust formation, and ultimately weaken both platform vitality and cross-tier conversion into paid services.

Generative artificial intelligence (genAI) represents a powerful technological intervention in this context. By functioning as a scalable and continuously available “super counselor,” genAI can dramatically improve responsiveness in public forums, alleviating human supply bottlenecks. Evidence from traditional content platforms suggests that AI integration can sustain user engagement when human contributions decline \citep{RN1365}. In principle, such improvements in responsiveness should expand patient participation and strengthen overall platform activity.

However, integrating genAI into OMHCs raises deeper concerns about professional participation and long-run sustainability. Research on single-tier knowledge platforms documents crowding-out effects, whereby AI reduces the perceived need for human input and dilutes the social recognition that motivates expert contribution \citep{RN1015,RN1209,RN1200,RN1215}. If similar dynamics arise in OMHCs, reduced counselor engagement in public forums could weaken patient–counselor interaction, undermine trust, and erode conversion into paid consultations—the core revenue source of the ecosystem. Moreover, because AI systems are trained on accumulated human knowledge, sustained professional contribution is essential for maintaining informational quality and relevance \citep{RN1247,10.1145/3571730,RN1240}.

Crucially, OMHCs differ from single-tier voluntary communities in their tiered structure. Public participation in the free forum directly shapes economic outcomes in the paid consultation tier \citep{RN307}. Counselors therefore contribute not only for intrinsic reasons, such as altruism or recognition, but also for strategic reasons tied to client acquisition \citep{RN1219}. In this environment, genAI integration does more than dilute intrinsic rewards. By increasing content density and improving responsiveness, AI reshapes both competitive visibility and the demand environment. Counselors who remain passive risk losing salience and downstream demand, whereas those who intensify participation may defend or expand their market share. At the same time, improved responsiveness can stimulate patient engagement, enlarging the pool of interaction opportunities and potentially generating complementarity between AI and human expertise rather than pure substitution.

These countervailing dynamics call for a careful examination of how they jointly determine equilibrium participation—an issue central to evaluating AI integration in tiered professional platforms. Our study takes a first step toward disentangling these mechanisms. Leveraging a quasi-natural experiment arising from the integration of a genAI-based conversational agent in one of the leading OMHCs in China, we examine how AI entry reshapes counselor participation. Using cross-platform difference-in-differences (CP-DID), calendar-year DID in time (CY-DID), and interrupted time series analysis (ITSA), we document a robust increase in counselors’ posting intensity following AI integration. At the same time, average per-post effort remains stable and per-post social recognition declines.

Mechanism analyses reveal that the aggregate increase in participation reflects interacting demand- and supply-side forces. On the demand side, AI significantly expands patient posting activity, with engagement partially reallocated from a nearby non-AI subforum. The resulting growth in threads enlarges counselors’ opportunity sets for contribution. On the supply side, counselors respond differently depending on their underlying incentives. Counselors with stronger economic orientation increase participation, consistent with intensified competitive incentives tied to paid-tier conversion, whereas counselors primarily driven by intrinsic motives reduce participation, consistent with recognition dilution and crowding-out. Importantly, these dynamics extend to the paid tier: counselors who remain active preserve or expand downstream demand, while inactive counselors experience economically meaningful declines in paid consultations.

Taken together, our findings show that genAI integration in tiered professional platforms does not inevitably erode human expertise. Instead, when public participation is linked to downstream economic outcomes, improved responsiveness and expanded engagement can activate strategic effort by counselors even as intrinsic social rewards are diluted. In this setting, genAI reshapes both the competitive and demand environment, allowing platforms to benefit simultaneously from automated scalability and sustained human participation. This coexistence is critical for OMHC sustainability: while genAI enhances access and responsiveness, counselors remain indispensable for delivering nuanced, emotionally sensitive support and for generating paid consultations that underpin platform revenue. Our results therefore highlight a governance perspective in which genAI and professional contributors function as complements rather than substitutes.

Our study contributes to three literatures. First, we extend research on OMHCs by moving beyond patient outcomes to examine how technological integration reshapes professional supply and strategic effort allocation \citep{RN273,RN16,RN101,RN305}. Second, we contribute to platform governance research by demonstrating that AI-based tools alter participation not only through cost changes but also through visibility-based competition and demand expansion in two-sided, tiered ecosystems. Third, we enrich the emerging literature on genAI in online communities \citep{RN1146, RN1248,RN1270,RN1224} by identifying institutional conditions under which AI complements rather than displaces professional participation.

The remainder of this paper proceeds as follows. Section~\ref{sec:lit} reviews the related literature. Section~\ref{sec:theorize} develops the theoretical framework. Section~\ref{sec:empirical} describes the empirical setting, data, and identification strategy. Section~\ref{sec:results} and \ref{sec:mechanism} present the main findings and mechanism analyses. Section~\ref{sec:discussion} concludes with theoretical and managerial implications.

\section{Related Literature}\label{sec:lit} 
\subsection{OMHCs as Tiered Two-Sided Platforms}

OMHCs connect patients seeking psychological support with licensed counselors through public Q\&A forums and paid private consultations. Prior research shows that these platforms can enhance psychological well-being and promote self-management \citep{RN1360,RN1359,RN273}. Their effectiveness, however, depends critically on timely and targeted interaction: merely observing others’ exchanges without receiving support can undermine outcomes for vulnerable users \citep{RN305}. Sustained responsiveness is therefore fundamental to OMHC functioning.

To ensure professionalization and financial sustainability, many OMHCs adopt a tiered structure in which a free public forum coexists with paid one-to-one services. In this setting, counselors’ public participation influences downstream demand: visible contributions signal competence and commitment, reduce patient uncertainty, and facilitate conversion to paid consultations \citep{RN16,RN101,RN307}. Public activity thus functions not only as knowledge provision but also as strategic reputation investment.

OMHCs therefore operate as tiered, two-sided platforms in which patient engagement and counselor effort are jointly determined \citep{RN792,RN1302}. Yet existing OMHC research has focused primarily on patient-side outcomes, leaving counselor participation and strategic effort allocation underexplored. In particular, little is known about how internal AI systems reshape the interaction between patient engagement, professional signaling, and competitive dynamics across tiers. Our study addresses this gap by examining how genAI integration alters participation incentives within a tiered OMHC ecosystem.

\subsection{Managing Online Content Platforms: Incentives, Competition, and Two-Sided Participation}

A central concern in platform research is how to sustain high-quality user-generated content. Existing studies identify mechanisms that stimulate contributions, including feedback systems \citep{RN1216}, identity disclosure \citep{RN1312}, social rewards \citep{RN1310}, and monetary incentives \citep{RN213,RN1311,RN1203}. These mechanisms strengthen intrinsic motivations or increase extrinsic returns.

Beyond explicit incentives, contribution behavior is shaped by competitive and attention environments. When contributors compete for limited visibility, greater content density can dilute recognition and weaken intrinsic motivation \citep{wu_huberman_2007,muchnik_2013}. Conversely, when rewards depend on relative standing or market share, intensified competition can increase equilibrium effort even without aggregate demand growth \citep{tullock1980,dixit1987}. On professional platforms, public contributions serve as observable signals that reduce consumer uncertainty and generate downstream demand \citep{RN1219,RN16,RN307,dellarocas_2003}. Technological shifts that alter visibility, substitution risk, or competitive pressure can therefore induce heterogeneous effort responses—crowding out some contributors while stimulating strategic effort from others.

Participation is also shaped by demand conditions. On two-sided platforms, contributor activity responds endogenously to user engagement: larger audiences and higher posting volumes expand contributors’ opportunity sets \citep{RN1302,RN792}. In the presence of genAI, improved responsiveness may stimulate such engagement rather than merely substitute for human participation. By lowering response latency and increasing the likelihood that patients receive timely replies, genAI can encourage more users to post and participate. The resulting growth in posting activity expands the pool of interaction opportunities for counselors. In professional environments where human expertise remains valuable for complex or emotionally sensitive cases, this engagement expansion can complement rather than replace human contributions. Thus, AI-induced demand growth may enlarge participation opportunities while simultaneously interacting with supply-side incentives.

Taken together, participation reflects the joint influence of intrinsic motives, competitive incentives, and demand dynamics. Prior research typically treats technological interventions as tools that directly affect contributor costs or rewards. We extend this literature by analyzing how genAI integration simultaneously reshapes competitive environments and expands demand within a tiered professional platform, producing heterogeneous supply responses.

\subsection{The Impact of GenAI on Online Communities}

GenAI systems autonomously produce content in response to natural language input \citep{RN1248}. Tools such as ChatGPT and GitHub Copilot have reshaped patterns of knowledge production and exchange \citep{RN909,RN1239}. Emerging evidence shows that external genAI tools can reduce question volumes, alter content composition, and affect labor demand in online communities \citep{RN1363,RN1270,RN1224,RN1244,RN1410}. These studies primarily document substitution effects in which AI replaces human participation.

Our study differs in two key respects. First, we examine internal genAI integration initiated by the platform rather than external tools that divert users away. Internal AI can complement engagement by improving responsiveness and retaining users within the ecosystem; for example, \cite{RN1365} show that banning AI-generated content reduces participation and content quality, highlighting AI’s role in sustaining demand.

Second, we focus on a tiered professional platform in which public participation affects downstream paid outcomes. Unlike single-tier voluntary communities \citep{RN1146,RN1365}, OMHCs link visibility in the free tier to revenue in the paid tier. In this environment, genAI integration reshapes both demand and competitive conditions. Improved responsiveness may expand patient engagement and enlarge participation opportunities, while increased content density and visibility competition may dilute intrinsic rewards and intensify strategic effort among economically motivated counselors. By integrating two-sided platform theory, signaling mechanisms, and heterogeneous incentive responses, our study advances understanding of how internal genAI reshapes participation in tiered online communities—not only through substitution or crowding-out, but also through demand-side complementarity and competition-driven effort.

\section{Theoretical Framework}\label{sec:theorize}

We develop a framework to analyze how integrating a genAI-based conversational agent reshapes counselors’ participation in OMHCs. OMHCs operate as tiered two-sided platforms in which public activity in the free forum affects economic outcomes in a downstream paid consultation market. In such environments, genAI integration alters not only the informational environment of the forum but also counselors’ incentives, competitive positioning, and the overall demand for interaction.

We identify three mechanisms that jointly shape equilibrium participation following genAI integration. First, AI-generated content may dilute visibility and reduce the marginal social return to contribution, generating intrinsic crowding-out. Second, because visibility in the public tier affects economic outcomes in the paid market, intensified competition may increase strategic effort among economically motivated counselors. Third, by improving responsiveness and stimulating patient engagement, genAI may expand the opportunity set for differentiated human input, creating demand-side complementarity between AI and counselors. These mechanisms operate simultaneously and interact to determine counselors’ equilibrium participation.

\begin{figure}
\FIGURE
{\includegraphics[scale=0.6]{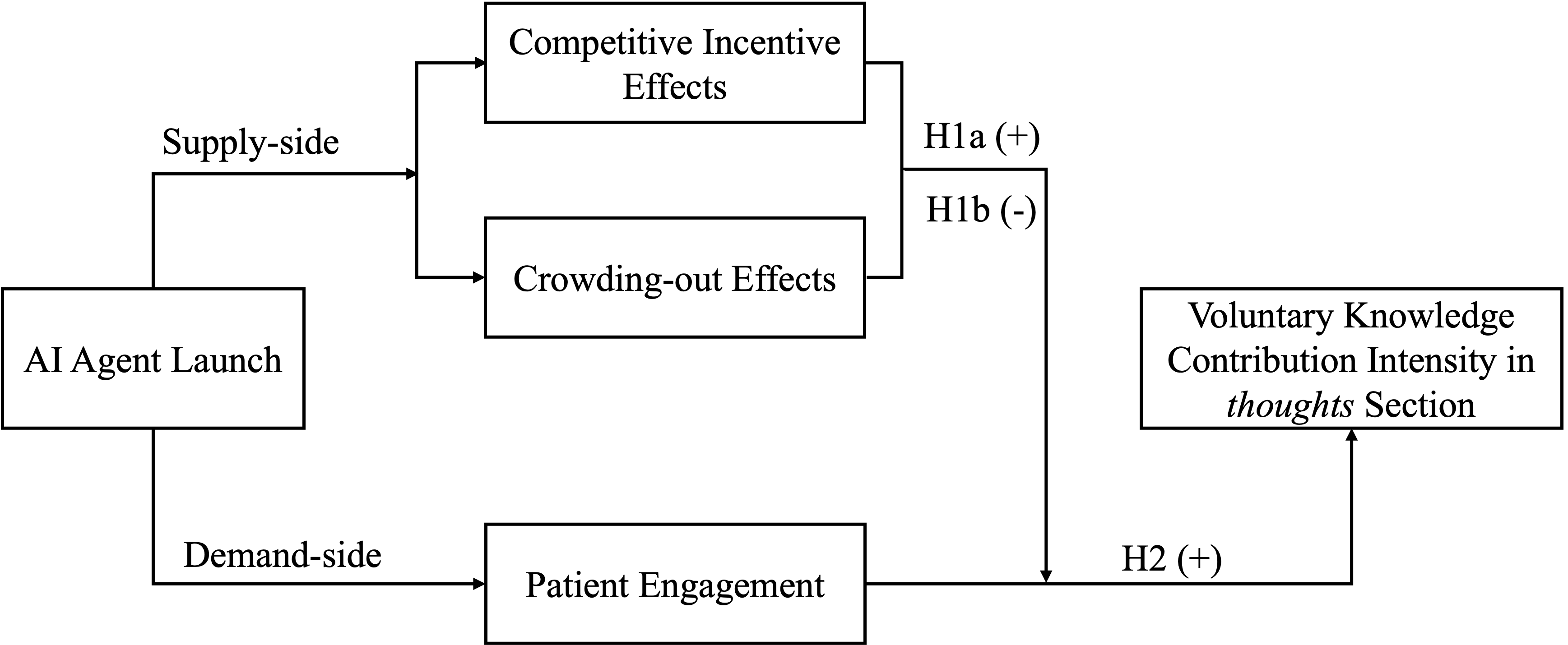}} 
{Theoretical Framework. \label{fig:hypo}}
{}
\end{figure}

Figure~\ref{fig:hypo} summarizes the conceptual framework.

\subsection{Supply-side Effects: Heterogeneous Incentive Responses}

We begin with the supply side. In tiered OMHCs, counselors contribute in a free public forum while simultaneously competing in a paid private consultation market. Participation therefore reflects a combination of intrinsic motivations and extrinsic economic incentives. The introduction of genAI may reshape these incentives in opposing directions.

\subsubsection{Intrinsic Crowding-out Effects}

In voluntary knowledge-sharing communities, participation is often driven by intrinsic motivations such as altruism, professional identity, competence validation, and social recognition \citep{RN1208,RN1307,RN1215}. For mental health counselors, public contributions function not only as informational responses but also as expressions of expertise and professional commitment.

GenAI integration may weaken these intrinsic drivers. First, genAI operates at scale and with high responsiveness, substantially increasing the volume of visible replies. As AI-generated content occupies greater attention space, the marginal visibility and distinctiveness of human responses decline \citep{wu_huberman_2007}. Because recognition and peer validation are central intrinsic rewards in online communities \citep{benabou_tirole_2006,RN1209,RN1200}, attention dilution reduces the expected social return to contribution, particularly in environments where early exposure shapes cumulative evaluation \citep{muchnik_2013}.

Second, when AI systems effectively address routine informational and emotional needs, the marginal validation derived from overlapping human responses diminishes. From a competence-based perspective \citep{deci_ryan_2000}, contributors derive utility from providing uniquely valuable input. If AI satisfies baseline needs, the perceived incremental impact of similar human contributions declines.

Under social exchange logic \citep{emerson_1976}, voluntary participation depends on whether expected social returns exceed marginal costs. Counselors incur time, cognitive, and emotional effort when responding to patient posts. If genAI reduces marginal visibility or distinctiveness while costs remain positive, intrinsically motivated counselors may rationally reduce participation.

Importantly, although genAI may lower the marginal cost of drafting by providing templates or reference answers, it can reduce the marginal return to contributions that are highly substitutable with AI output. When recognition depends on relative distinctiveness, redundant content generates limited visibility-based rewards. In equilibrium, contributions with low marginal distinctiveness are therefore less likely to be sustained.

\textbf{Channel 1a (Intrinsic Crowding-out).} \textit{GenAI integration reduces participation among counselors whose motivation is primarily intrinsic.}

\subsubsection{Extrinsic Competitive Incentive Effects}

At the same time, OMHCs embed economic incentives that are absent in purely voluntary communities. Public participation influences downstream paid demand: visible contributions signal competence and reduce patient uncertainty, facilitating conversion to private consultations \citep{RN1219,RN16,RN307,dellarocas_2003}. Public activity therefore serves as a strategic investment in reputation and market position.

GenAI integration reshapes this competitive environment. By increasing content density and accelerating responsiveness, it reduces the baseline visibility of individual counselors. Moreover, by providing low-cost informational and emotional support, genAI introduces substitution risk for paid services \citep{RN1411}. Both forces threaten expected economic returns if counselors remain passive.

However, when outcomes depend on relative visibility rather than absolute demand, intensified competition can increase equilibrium effort \citep{tullock1980,dixit1987}. In such environments, effort determines relative market share. As competitive pressure rises, economically motivated counselors may increase participation in order to defend salience, differentiate expertise, and preserve conversion opportunities. Public contribution thus becomes a strategic response to AI-induced competition.

\textbf{Channel 1b (Competitive Incentive Intensification).} \textit{GenAI integration increases participation among counselors whose behavior is primarily driven by extrinsic economic incentives.}

\subsection{Demand-side Effects: Engagement Expansion and Complementarity}

Beyond supply-side incentives, genAI integration also reshapes participation through demand dynamics.

Patients’ willingness to post depends heavily on platform responsiveness—the likelihood and speed of receiving replies \citep{lu2017core,RN1213,RN1358}. When expected response probability is low, participation declines. By providing real-time, scalable, and continuous replies, genAI substantially increases response likelihood and reduces waiting time \citep{RN1365}. From a social exchange perspective \citep{emerson_1976}, this raises the expected return to posting. Because posting costs remain largely unchanged, improved responsiveness increases net expected utility \citep{lu2017core}.

As responsiveness improves, patient engagement expands along both extensive and intensive margins: more patients participate, and existing patients post more frequently and with more nuanced concerns. The resulting growth in threads enlarges the overall interaction space.

Crucially, this expansion does not imply pure substitution between AI and human responses. While genAI may satisfy standardized informational needs, heterogeneous, complex, or emotionally sensitive cases often remain only partially addressed. Because counselor participation is largely reactive to patient posts, increased engagement expands the set of cases in which differentiated human expertise remains valuable. In this sense, AI-induced demand growth can complement rather than replace human participation.

\textbf{Channel 2 (Demand-side Complementarity).} \textit{GenAI integration expands patient engagement, thereby enlarging counselors’ opportunity sets and increasing participation.}

\subsection{Net Equilibrium Implications}

These mechanisms operate simultaneously following genAI integration. Intrinsic crowding-out predicts reduced participation among recognition-driven contributors, whereas competitive incentive intensification predicts increased effort among economically motivated counselors. At the same time, demand-side complementarity expands the overall opportunity set for contribution.

Importantly, demand expansion and supply-side responses are not independent. Increased patient engagement generates additional participation opportunities, but the realization of these opportunities depends on counselors’ discretionary effort. In a tiered professional environment where public visibility affects downstream economic outcomes, expanded engagement can therefore activate strategic responses among counselors who seek to maintain or enhance their market position.

Because visibility in the public tier directly affects economic outcomes in the paid tier, the tiered structure structurally amplifies competitive incentives. Combined with expanded engagement opportunities, this amplification may outweigh intrinsic crowding-out effects.

Accordingly, although the net effect is theoretically ambiguous ex ante, the institutional structure of OMHCs suggests a positive equilibrium response in counselors’ participation.

\textbf{Hypothesis.} \textit{GenAI integration increases counselors’ overall participation intensity.}

\section{Research Context and Empirical Strategy}\label{sec:empirical}
\subsection{Research Context}
We study one of the largest OMHCs in China, which connects over 43 million registered patients with approximately 42,000 licensed counselors. The platform operates under a tiered service structure: counselors may (1) participate in a public Q\&A forum by responding to patient posts for free, and (2) offer paid one-to-one private consultations. The public forum functions as an entry point for patient interaction, while paid consultations constitute the platform’s main revenue source.

\begin{figure}
\FIGURE
{\includegraphics[scale=0.5]{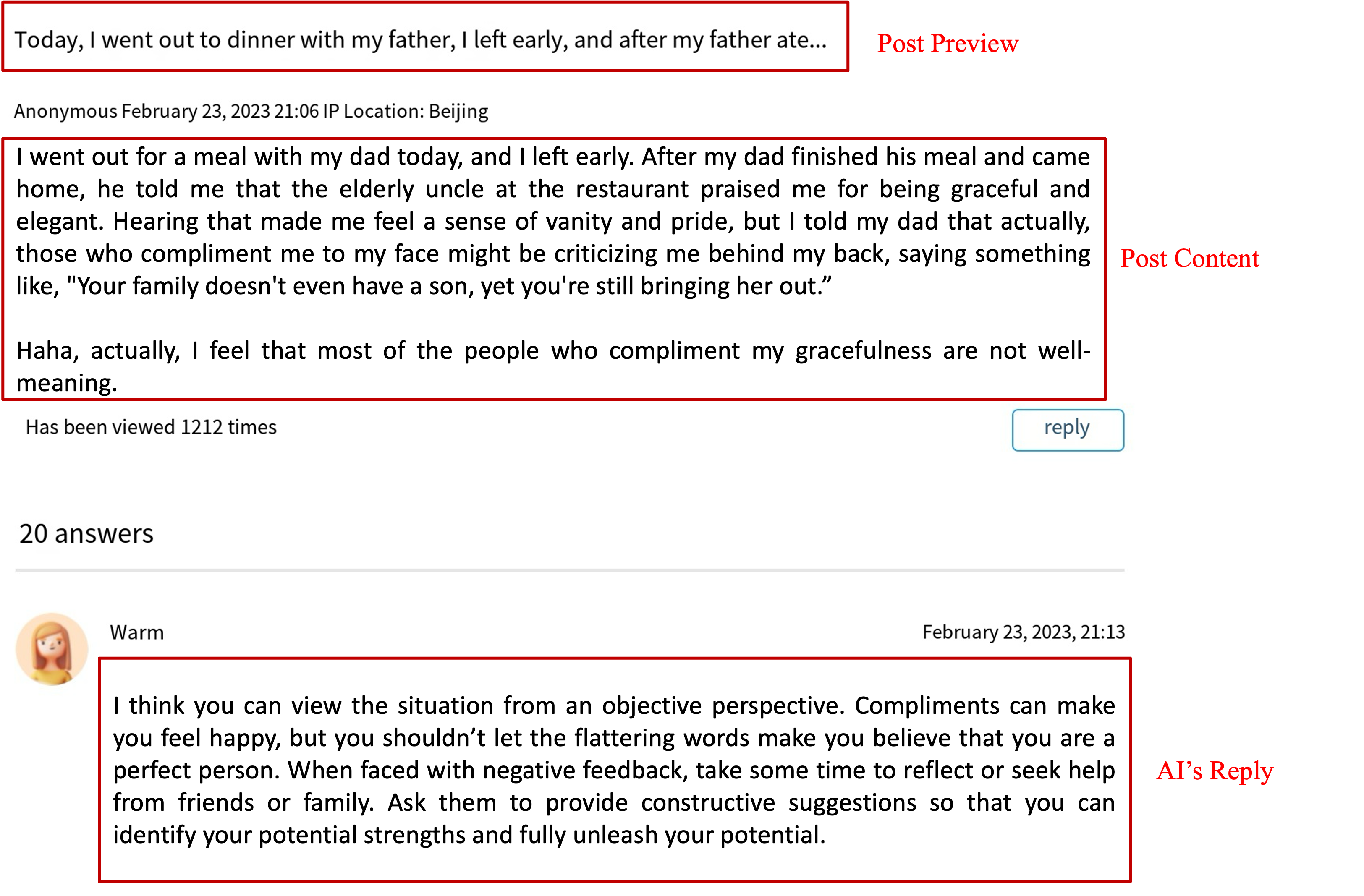}} 
{A Conversation Between AI and a Patient. \label{fig:AI_screen}}
{}
\end{figure}

In February 2023, the platform introduced a genAI–based conversational agent designed to provide complementary counseling support. The AI agent acts as a regular community participant, automatically replying to patient posts in one of the two main subforums—\textit{Thoughts}. Equipped with natural language understanding and empathy modeling, the agent can provide both informational support (e.g., problem analysis and practical suggestions) and emotional support (e.g., expressions of empathy and encouragement). Unlike rule-based bots studied in earlier work \citep{RN1162,RN452,RN1280}, the agent engages in two-way, back-and-forth interactions, dynamically generating new replies as patients continue responding. This design allows the AI to emulate the conversational depth and continuity of a human counselor. Figure \ref{fig:AI_screen} presents an exemplar conversation between the AI agent and a patient. 

Importantly, the genAI-based agent was launched only in the \textit{Thoughts} subforum, not in the \textit{Answers} subforum. Both subforums are free for patients and counselors to use, but they differ in structure and tone. Posts in \textit{Thoughts} require a title and have no length restrictions, fostering more casual, open-ended exchanges. In contrast, posts in \textit{Answers} must exceed twenty words and tend to involve structured, content-focused Q\&A interactions. Because \textit{Thoughts} is more conversational and tolerant of stylistic variation, it was selected as the pilot site for genAI deployment, minimizing user dissatisfaction risk while maximizing engagement potential. This platform-wide deployment creates a quasi-natural experimental setting to identify the causal effects of genAI integration on counselor behavior.

\subsection{Empirical Strategy}

We leverage the introduction of the genAI conversational agent—an \textit{exogenous and time-bound shock}—using two complementary identification strategies: a \textit{DID} framework and a \textit{ITSA} design. Together, these approaches enable robust causal inference on how genAI integration affects counselors’ participation behaviors.

\subsubsection{DID}

Ideally, a DID framework compares counselor activity before and after the AI integration against suitable control groups. Because all counselors were simultaneously exposed to the treatment, we construct two alternative counterfactuals to ensure credible identification:

\begin{enumerate}
    \item \textbf{CP-DID.} We use another major OMHC in China—where no AI deployment occurred during the same period—as the control group. Compared to the ``unaffected'' \textit{Answers} subforum, an entirely different platform provides a more remote and naturally isolated comparison, minimizing interference concerns. To improve comparability, we perform \textit{propensity score matching (PSM)} based on pre-treatment counselor characteristics, including service domains, number of patients, average ratings, qualifications, and counseling styles. We apply one-to-one nearest-neighbor matching with a caliper of 0.05 and no replacement, yielding 492 treated and control counselors. Balance diagnostics confirm strong post-matching comparability \citep{RN1242,RN907}.\footnote{To assess the quality of matching, we depict the distribution of propensity score for both treatment and control group counselors before and after the matching in Figure \ref{fig:cross_psm} in Appendix \ref{sec:psm_appendix} , following \citep{RN1242,RN907}. Figure \ref{fig:cross_psm} shows substantially consistent distributions for the treatment and control group, indicating that the two groups are well matched.}
    
    \item \textbf{CY-DID.} As a within-platform benchmark, we compare each counselor’s activity in 2023 (the AI integration year) with their own outcomes in the same calendar months of 2022. This design controls for unobserved individual heterogeneity and platform-level shocks, and more closely satisfies the \textit{Stable Unit Treatment Value Assumption (SUTVA)}, though it may be somewhat more sensitive to seasonal variation.
\end{enumerate}

We estimate the following baseline model:
\begin{equation}
Outcome_{itd} = \beta_1 (Treat_d \times PostLaunch_t) 
 + \mu_t + \lambda_{id} + \varepsilon_{itd},
\label{eq:DID_doc}
\end{equation}
where \(Outcome_{itd}\) represents the logarithm of counselor \(i\)’s participation intensity in week \(t\). \(PostLaunch_t\) equals 1 after February 17, 2023, and \(Treat_d\) indicates treated counselors (platform-level in CP-DID; year-level in CY-DID). The coefficient \(\beta_1\) captures the causal effect of AI integration. Counselor–year fixed effects \((\lambda_{id})\) absorb time-invariant heterogeneity, while week fixed effects \((\mu_t)\) control for common shocks across counselors. Standard errors are clustered at the counselor level.

\subsubsection{ITSA}

To further corroborate our findings, we employ a ITSA design exploiting the sharp timing of the AI integration. The launch date was predetermined by platform administrators and unrelated to counselor behavior, creating a clean temporal cutoff. We estimate:
\begin{align}
Outcome_{it} = \beta_1 PostLaunch_t
+ f(RelTime_t) 
+ PostLaunch_t \times f(RelTime_t) 
+ \lambda_i + \varepsilon_{it},
\label{eq:ITSA_doc}
\end{align}
where \(RelTime_t\) measures the number of weeks relative to the launch date. The flexible polynomial \(f(\cdot)\), specified as first order following \cite{RN1199}, captures smooth temporal trends on either side of the cutoff. The coefficient \(\beta_1\) identifies the \textit{immediate behavioral shift} in counselor participation triggered by the AI integration.

Importantly, this ITSA specification estimates an \textit{average treatment effect at the time of treatment}, capturing the instantaneous change upon integration rather than the averaged post-period effects estimated by DID. As such, while magnitudes may differ across methods, we expect both estimates to align directionally under our theoretical predictions.

\subsection{Data and Descriptive Findings}

From the focal OMHC, we collect a panel data for a random sample of \textit{1,663 active counselors} who participated in either the public Q\&A forum or the paid consultation services during the observation period. For the CP-DID design, we supplement a random sample of 533 counselors from the control OMHC—another leading OMHC in China that did not integrate any AI system during the same period. 

For each counselor, we collect three types of time-series data at weekly level:

\begin{itemize}
    \item \textbf{Public participation data:} reply texts, and the number of likes for all responses made by counselors in both the \textit{Thoughts} and \textit{Answers} sections;
    \item \textbf{Paid service data:}  patient identifiers, session lengths, and patient ratings for one-to-one consultations;
    \item \textbf{Profile data:} counselor-level attributes including service domains, number of credible qualifications, and counseling style.
\end{itemize}

We also collect all AI-generated responses, including timestamps, post content, and the number of likes received. 

For the CP-DID and ITSA designs, we collect the data for 47 weeks in 2023, spanning the 10 weeks before and 36 weeks after the AI launch date. For the CY-DID design, we additionally collected 47 weeks lagged by one year (the same calendar weeks in 2022) for the same set of counselors. Additionally, for the focal OMHC, we collected data on all replies made by counselors in both the \textit{thoughts} and \textit{answers} sections. The later allows us to examine the spillovers.

\begin{figure}
\FIGURE
{\includegraphics[scale=0.45]{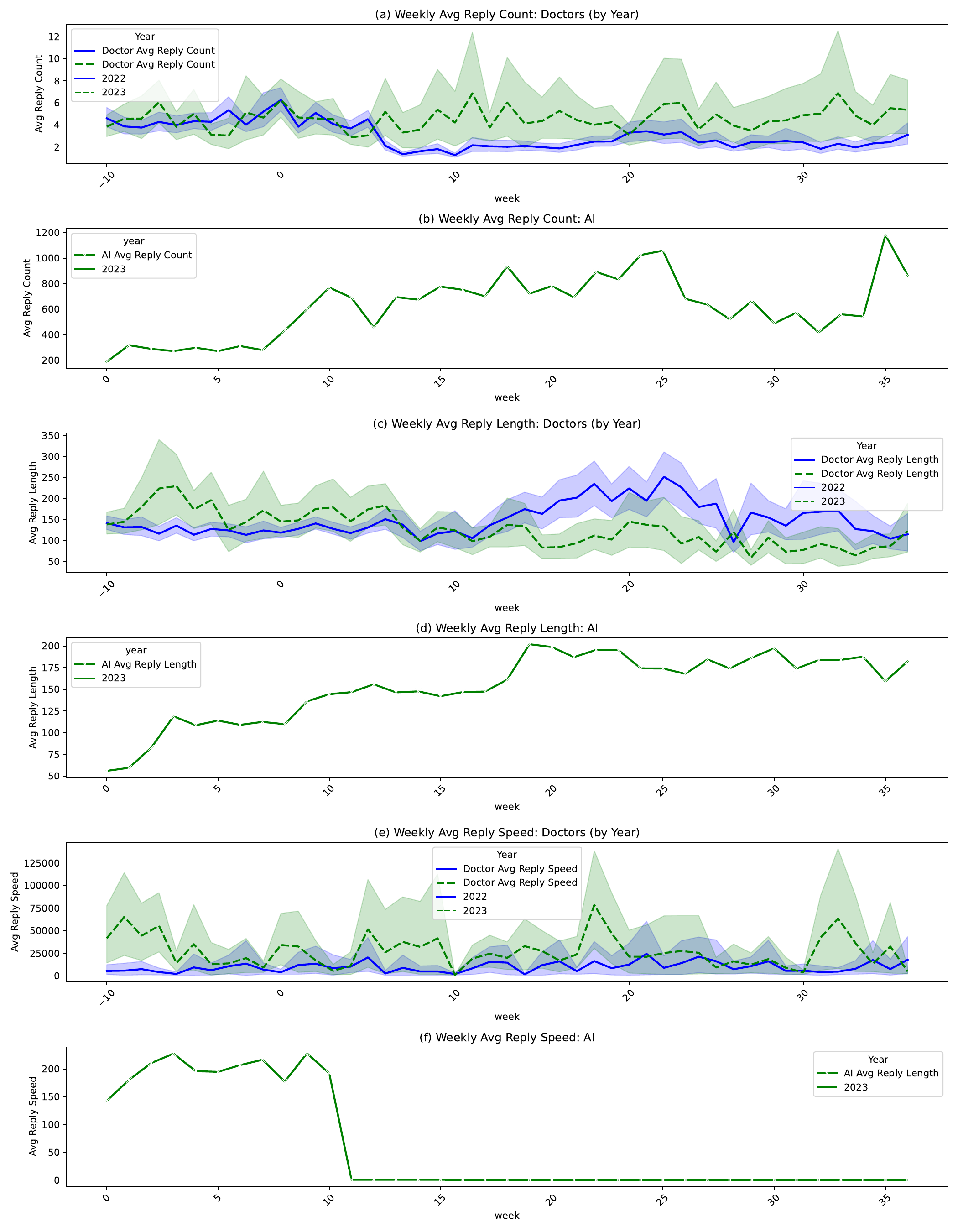}} 
{Weekly Participation in the \textit{Thoughts} Sub-section of counselors and AI. \label{fig:model_free}}
{The green lines represent the statistics for 2023, while the blue lines represent those for 2022. Since the AI was introduced to the OMHC after week 0 in 2023, no data is available for it in 2022 or before week 0 in 2023.}
\end{figure}

Figure~\ref{fig:model_free} presents descriptive trends in response behavior. The genAI agent exhibits a substantially higher response capacity, generating nearly 100 times more replies per week than an average counselor and achieving near real-time responsiveness (under 200 minutes) by week 11. Despite this difference in scale, the average reply lengths are comparable—100–200 words for the AI agent versus 100–250 words for counselors—indicating a high degree of content substitutability. Table~\ref{tab:DS} summarizes the variable definitions and descriptive statistics used in the analysis.

\begin{table}
\TABLE
{Definition and Summary Statistics of Variables.\label{tab:DS}}
{\begin{tabular}{@{}lp{6.5cm}ccccc@{}}
\hline \up
Variable          & Definition                                                    & N        & Mean     & Std.     & Min     & Max      \\ \hline
 \multicolumn{7}{l}{\textbf{Contribution Efforts in AI Invovled Section}}\\ 
$ThoughtIntensity_{it}$& The number of times that the counselor $i$ replies to posts in the \textit{thoughts} section in week $t$.& 156,322 & 0.077 & 1.111 & 0.000 & 78.000 \\ 
$ThoughtLength_{it}$& The average number of words (excluding stopwords) in counselor $i$’s replies to posts in the \textit{thoughts} section in week $t$.& 2,173 &  89.799 &   96.338 & 1.000 &   963.500 \\
$ThoughtLikes_{it}$& The average number of likes that counselor $i$ received by posts in the \textit{thoughts} section in week $t$.&   2,173& 0.162& 0.428& 0.000 & 10.000 \\
 \multicolumn{7}{l}{\textbf{Contribution Efforts in non-AI Invovled Section}}\\
 $AnswerIntensity_{it}$& The number of times that the counselor $i$ replies to posts in the \textit{answers} section in week $t$. & 156,322 & 0.322 & 3.259 & 0.000 &292.000 \\
 \multicolumn{7}{l}{\textbf{Demand for Fee-based Service}}\\ 
$Demand_{it}$            & The number of transactions for the counselor's paid private services.& 156,322 &  1.250 & 4.513 & 0.000 & 166.000 \\
 \multicolumn{7}{l}{\textbf{Other Variables}}\\
$TPostVolume_{t}$& The total number of posts that patients post in the \textit{thoughts} section in week $t$.&  156,322 & 560.415&  243.973& 205.000& 1819.000\\ 
$APostVolume_{t}$& The total number of posts that patients post in the \textit{answers} section in week $t$.&  156,322 &  613.532& 465.079& 116.000& 1984.000 \down\\ \hline
\end{tabular}}{}
\footnotesize
        \textit{Note}: The quality measure for thoughts, such as likes and reply length, doesn't exist when counselors do not respond in the forum or provide paid private services. Therefore, there are fewer than 156,322 observations for these specific variables. 
\end{table}

\section{Main Results}\label{sec:results}
\subsection{Baseline Effects in the AI-Involved \textit{Thoughts} Subforum}\label{sec:main_thoughts}

Table~\ref{tab:doc_lab} presents baseline estimates from three complementary identification strategies: CP-DID, CY-DID, and ITSA. Across all specifications, the post-integration effect on counselors’ weekly posting volume (\textit{ThoughtIntensity}) is positive and statistically significant (Columns~(1)–(3)). The consistency across designs indicates that the introduction of the genAI agent is associated with a robust increase in counselor participation in the AI-enabled \textit{Thoughts} subforum.

This aggregate increase is notable because AI simultaneously expands total answer supply and intensifies competition for attention. Rather than contracting in the presence of a scalable “super-supplier,” counselors increase their posting intensity. This pattern is consistent with the theoretical framework, suggesting that demand expansion and competitive effort responses dominate any pure intrinsic crowding-out pressure at the aggregate level.

We next examine whether the increase reflects deeper per-post investment. Columns~(4)–(6) show no statistically significant change in average reply length (\textit{ThoughtLength}), indicating stable per-post effort following AI integration. Adjustment therefore occurs primarily along the extensive margin—counselors post more frequently rather than investing more heavily in each individual response.

In contrast, Columns~(7)–(9) reveal a statistically significant decline in likes per reply (\textit{ThoughtLikes}). This decline is consistent with attention dilution in a denser content environment created by abundant AI-generated responses. Lower per-post recognition aligns with the intrinsic crowding-out mechanism, as the marginal social return to each contribution falls when visibility becomes more contested.

Importantly, the coexistence of reduced recognition and increased posting frequency provides indirect evidence of intensified competitive incentives. In a tiered professional setting where public visibility affects downstream paid outcomes, counselors compete for relative salience rather than absolute demand. When competition for attention intensifies, maintaining or defending market position requires incremental participation. Increased posting can therefore be interpreted as a strategic response to heightened competition. Absent such competitive incentives, declining recognition—and the associated weakening of intrinsic motivation—would instead predict reduced participation.

\begin{table}
\TABLE
{The Impact of AI Launch on counselors' Voluntary Knowledge Contribution Effort\label{tab:doc_lab}}
{\begin{tabular}{@{}lccccccccc@{}}
\hline \up
& \multicolumn{3}{c}{\textit{ThoughtIntensity}}& \multicolumn{3}{c}{\textit{ThoughtLength}}& \multicolumn{3}{c}{\textit{ThoughtLikes}}\\
& (1)&(2)& (3)  & (4)&(5)& (6)&(7)&(8)&(9)\\ 
& CP-DID&CY-DID& ITSA  & CP-DID&CY-DID& ITSA  & CP-DID&CY-DID& ITSA  \\ \hline \up
\textit{Treat×PostLaunch}& 0.006*
&0.025***& & 0.052
&0.113 & 
& -0.125**
&-0.102*
&
\\
& (0.003)
&(0.003)& & (0.103)
&(0.107) & 
& (0.056)
&(0.052)
&
\\
\textit{PostLaunch}& && 0.009***&  && 0.049
&  &&-0.148*
\\
& &&  (0.003)&  && (0.137)
&  &&(0.080)
\\
Constant & 0.015***
&0.012***&  0.007***& 4.753***
&4.078*** & 4.134***
& 0.679***
&0.135***
&0.284***
\\
& (0.001)
&(0.001)&  (0.003)& (0.049)
&(0.023) & (0.130)
& (0.026)
&(0.011)
&(0.072)
\\
Observations & 46,248
&156,322&  78,161 & 1,093
&1,891 & 424& 1,093
&1,891
&424
\\
R-squared & 0.391
&0.210& 0.282 & 0.838
&0.728 & 0.753& 0.876
&0.375
&0.322
\\
Weekly FE & Yes &Yes & -& Yes &Yes & -& Yes &Yes &-\\
Year-Counselor FE/Counselor FE& Yes &Yes & Yes & Yes &Yes& Yes & Yes &Yes&Yes  \down\\ \hline
\end{tabular}}{Robust standard errors in parentheses. *** p\textless{}0.01, ** p\textless{}0.05, * p\textless{}0.1}
\end{table}

\subsection{Pre-treatment Trends and Dynamic 
Patterns}\label{sec:validity_dynamic}

The causal interpretation of the DID estimates relies on a parallel-trends assumption in the absence of treatment. We assess this assumption for the baseline CY-DID design using an event-study specification with relative-week indicators, following \cite{RN1282} and \cite{RN904}:
\begin{align}
Outcome_{itd}= 
\beta_0  + \sum_{j}\gamma_{j}\cdot(Pre_{t}(j)\times Treat_d)
+ \sum_{k}\lambda_{k}\cdot(Post_{t}(k)\times Treat_d) 
+\mu_t + \lambda_{id} + \varepsilon_{itd},
\label{eq:RTM}
\end{align}
where \(Pre_t(j)\) equals one if week \(t\) is \(j\) weeks before the integration date (February~17) in the relevant year, and \(Post_t(k)\) equals one if week \(t\) is \(k\) weeks after. We bin all weeks that are ten or more weeks away from the event into \(Pre_t(10)\) and \(Post_t(10)\). The indicator for one week prior to the launch, \(Pre_t(1)\), is omitted as the reference period.

Under the identifying assumption, the pre-treatment coefficients \(\gamma_j\) should be statistically indistinguishable from zero. As shown in Table~\ref{tab:main_rtm}, all pre-treatment estimates are small and statistically insignificant, supporting the parallel-trends assumption and alleviating concerns about differential seasonality or pre-existing trends.

\begin{figure}
\FIGURE
{
\subcaptionbox{ThoughtIntensity\label{fig:cross_rtm1}}
{\includegraphics[width=0.3\textwidth]{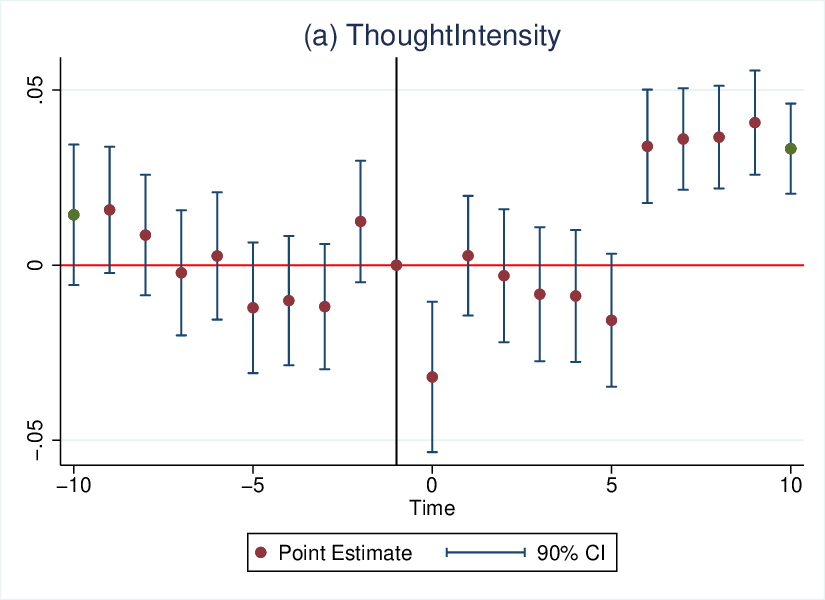}}
\hfill\subcaptionbox{ThoughtLength\label{fig:cross_rtm2}}
{\includegraphics[width=0.3\textwidth]{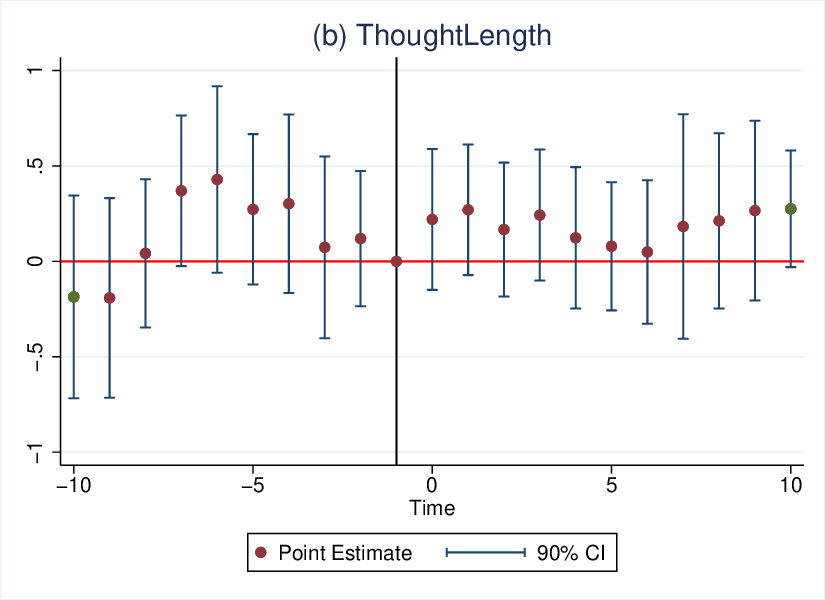}}
\hfill\subcaptionbox{ThoughtLikes\label{fig:cross_rtm3}}
{\includegraphics[width=0.3\textwidth]{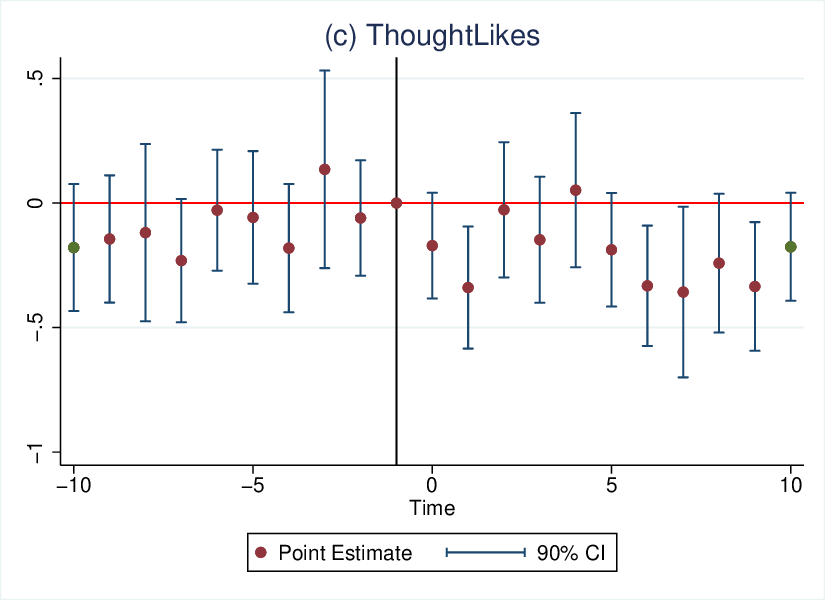}}
}
{
Dynamic Effects of the GenAI-based Conversational Agent Launch
\label{fig:cross_main_rtm}}
{
}
\end{figure}

The event-study estimates also reveal informative dynamic patterns. Figure~\ref{fig:cross_main_rtm} shows that counselors’ posting intensity increases gradually, with a noticeable response emerging several weeks after the integration date. This lagged adjustment is consistent with a learning and adaptation process rather than a purely mechanical response to AI entry. 

In contrast, the trajectory of likes exhibits an immediate decline around the launch and remains persistently lower thereafter, consistent with rapid dilution of user attention following the introduction of pervasive AI-generated responses. Together, these dynamics reinforce the interpretation that genAI integration reshapes both participation incentives and recognition mechanisms in the community.

\begin{table}
\TABLE
{The Dynamic Effects of the GenAI-based Conversational Agent Launch\label{tab:main_rtm}}
{\begin{tabular}{llll}
\hline \up
                  & (1)              & (2)           & (3)          \\
                  & ThoughtIntensity & ThoughtLength & ThoughtLikes \\ \hline
Pre10             & 0.014            & -0.186        & -0.179       \\
                  & (0.012)          & (0.323)       & (0.155)      \\
Pre9              & 0.016            & -0.191        & -0.144       \\
                  & (0.011)          & (0.318)       & (0.155)      \\
Pre8              & 0.009            & 0.042         & -0.119       \\
                  & (0.010)          & (0.236)       & (0.216)      \\
Pre7              & -0.002           & 0.370         & -0.231       \\
                  & (0.011)          & (0.240)       & (0.150)      \\
Pre6              & 0.003            & 0.429         & -0.029       \\
                  & (0.011)          & (0.297)       & (0.147)      \\
Pre5              & -0.012           & 0.273         & -0.058       \\
                  & (0.011)          & (0.239)       & (0.162)      \\
Pre4              & -0.010           & 0.302         & -0.181       \\
                  & (0.011)          & (0.284)       & (0.156)      \\
Pre3              & -0.012           & 0.074         & 0.135        \\
                  & (0.011)          & (0.290)       & (0.241)      \\
Pre2              & 0.012            & 0.120         & -0.060       \\
                  & (0.011)          & (0.215)       & (0.141)      \\
Post0             & -0.032**         & 0.220         & -0.171       \\
                  & (0.013)          & (0.224)       & (0.129)      \\
Post1             & 0.003            & 0.270         & -0.339**     \\
                  & (0.010)          & (0.208)       & (0.149)      \\
Post2             & -0.003           & 0.167         & -0.027       \\
                  & (0.012)          & (0.213)       & (0.165)      \\
Post3             & -0.008           & 0.243         & -0.148       \\
                  & (0.012)          & (0.209)       & (0.154)      \\
Post4             & -0.009           & 0.123         & 0.052        \\
                  & (0.011)          & (0.225)       & (0.188)      \\
Post5             & -0.016           & 0.079         & -0.188       \\
                  & (0.012)          & (0.204)       & (0.139)      \\
Post6             & 0.034***         & 0.049         & -0.332**     \\
                  & (0.010)          & (0.228)       & (0.147)      \\
Post7             & 0.036***         & 0.183         & -0.357*      \\
                  & (0.009)          & (0.358)       & (0.208)      \\
Post8             & 0.037***         & 0.212         & -0.241       \\
                  & (0.009)          & (0.279)       & (0.169)      \\
Post9             & 0.041***         & 0.266         & -0.335**     \\
                  & (0.009)          & (0.286)       & (0.157)      \\
Post10            & 0.033***         & 0.276         & -0.176       \\
                  & (0.008)          & (0.186)       & (0.132)      \\
Constant          & 0.011***         & 4.050***      & 0.161***     \\
                  & (0.004)          & (0.040)       & (0.030)      \\
Observations      & 156,322          & 1,891         & 1,810        \\
R-squared         & 0.211            & 0.730         & 0.411        \\
Control Variables & No               & No            & No           \\
Year-Counselor FE & Yes              & Yes           & Yes          \\
Weekly FE         & Yes              & Yes           & Yes         \down \\ \hline 
\end{tabular}}{Robust standard errors in parentheses. *** p\textless{}0.01, ** p\textless{}0.05, * p\textless{}0.1}
\end{table}

\section{Mechanism Analysis}\label{sec:mechanism}

The baseline findings—higher posting intensity, stable reply length, and reduced per-post likes—suggest a structured equilibrium adjustment consistent with our theoretical framework. In a tiered two-sided setting, genAI integration can simultaneously (i) dilute recognition and weaken intrinsic returns, (ii) intensify competitive incentives tied to downstream paid services, and (iii) expand patient engagement, thereby enlarging counselors’ opportunity sets. Importantly, these forces operate jointly rather than independently: demand expansion reshapes the competitive environment faced by counselors, while supply-side responses determine how newly created participation opportunities are realized. We therefore examine the empirical presence of these mechanisms, beginning with the demand-side channel.

\subsection{Demand-side Channel: Patient Engagement Expansion and Complementarity}\label{sec:mechanism_demand}

Our framework predicts that genAI improves platform responsiveness, thereby stimulating patient engagement. Crucially, in a two-sided professional ecosystem, this demand expansion need not substitute for human supply. Instead, AI-induced engagement growth can enlarge the set of heterogeneous and nuanced cases in which differentiated counselor input remains valuable.

We operationalize patient demand using weekly patient posting volume in the \textit{Thoughts} subforum (\(TPostVolume_{td}\)). Column (1) of Table~\ref{tab:inten_mechanism} presents CY-DID estimates with \(TPostVolume_{td}\) as the dependent variable. The coefficient on \textit{Treat×PostLaunch} is positive and statistically significant, indicating that genAI integration substantially increases patient posting activity.

One potential concern is that increased counselor participation may itself stimulate patient posting. To address this possibility, Column (2) includes total counselor participation as a control. The post-integration coefficient remains positive, statistically significant, and economically large. This pattern suggests that the increase in patient posts is not merely induced by counselor activity, but is primarily driven by the enhanced responsiveness and availability introduced by genAI.

Next, we examine whether expanded patient engagement translates into greater counselor participation through complementarity. To do so, we introduce \(\log(TPostVolume_{td}+1)\) as a mediator into the main participation specification:

\begin{align}
Outcome_{itd} =
\beta_1 (Treat_d \times PostLaunch_t)
+ \beta_2 \log(TPostVolume_{td}+1)
+\mu_t + \lambda_{id} + \varepsilon_{itd}.
\label{eq:inten_mechanism1}
\end{align}

Column (3) shows that patient posting volume is strongly and positively associated with counselor participation. Once patient demand is controlled for, the post-integration indicator becomes statistically insignificant. This attenuation suggests that the aggregate increase in counselor participation operates largely through demand expansion.

Importantly, this mediation does not reflect mechanical substitution between AI and human supply. Instead, it reflects complementarity between AI and human counselors. While genAI generates initial responses and improves baseline responsiveness, the resulting increase in total threads expands the pool of cases in which differentiated human expertise remains valuable. Because counselor participation is largely reactive to patient posts, growth in the volume and diversity of threads directly enlarges the opportunity set for human contribution.

At the same time, the demand-side mechanism cannot be cleanly separated from supply-side responses. The insignificant treatment effect after controlling for patient demand should not be interpreted as evidence that supply-side incentives are absent. Rather, demand expansion and supply adjustments are jointly determined. More rigorously, the results suggest that the increase in counselor participation following AI entry is proportional to the increase in patient posts. Because patient posts represent participation opportunities, the observed rise in activity implies that counselors actively respond to these opportunities rather than passively withdrawing. In a tiered professional environment, such behavior is more consistent with intensified competitive incentives than with passive crowding-out due to weakened intrinsic returns.

Column (4), which replaces the binary treatment indicator with the continuous AI reply measure (\textit{AIVolume}), yields consistent results. Together, these findings provide strong evidence that genAI-induced demand expansion is a central mechanism underlying the positive net participation effect. In a tiered professional setting, improved responsiveness generates complementary human participation opportunities rather than purely crowding them out.

\begin{table} 
\TABLE
{Effects of Patient Engagement and AI Competitive Incentives on Counselors' Voluntary Knowledge Contribution Intensity in \textit{thoughts} Section \label{tab:inten_mechanism}
}
{\begin{tabular}{@{}lllcc}
\hline \up
&   (1) &(2)   &(3)& (4)\\
&   TPostVolume&TPostVolume&ThoughtIntensity & ThoughtIntensity \\ \hline \up
\textit{Treat×PostLaunch}&   0.734***&0.452***&0.002& \\
                              &   (0.090)&(0.134)&(0.004)& \\
 \textit{AIVolume}&   &&& 0.001\\
 &   &&& (0.001)\\
 \textit{TPostVolume}&   &&0.031***& 0.029***\\
 &   &&(0.003)& (0.003)\\
 \textit{TThoughtIntensity}&   &0.276***&& \\
 &   &(0.063)&& \\
Constant                      &   5.959***&4.617***&-0.174***& -0.160***\\
                              &   (0.043)&(0.313)&(0.018)& (0.019)\\
Observations                  &   94&94&156,322& 156,322\\
R-squared                     &   0.826&0.884&0.210& 0.210\\
Year-Counselor FE&   No&No&Yes                      & Yes                      \\
 Yearly FE&   Yes                        &Yes                         &No& No\\
 Weekly FE&   Yes                        &Yes                         &Yes                      & Yes                      \\ \hline
\end{tabular}}{Robust standard errors in parentheses. *** p\textless{}0.01, ** p\textless{}0.05, * p\textless{}0.1}
\end{table}

\subsection{Supply-side Channel: Offsetting Incentives and a Boundary Condition}\label{sec:mechanism_supply}

On the supply side, genAI integration can simultaneously weaken intrinsic recognition-based rewards while intensifying competitive incentives tied to downstream paid services. Because these forces operate in opposite directions, they may offset each other in aggregate regressions and are empirically difficult to separate. Accordingly, while the previous results suggest that competitive responses dominate at the aggregate level, they do not imply the absence of intrinsic crowding-out.

However, these mechanisms may manifest differently across heterogeneous counselors. Some counselors are primarily intrinsically motivated, deriving utility from helping patients and receiving social recognition. Others are more economically oriented and view public participation as a strategic channel for attracting paid consultations. This heterogeneity allows us to isolate the presence of each mechanism that may otherwise be masked in aggregate patterns.

To examine these mechanisms, we classify counselors based on pre-launch behavior as a proxy for motivational orientation. Intrinsic-dominant counselors are those who actively contributed in the public forum but generated no paid consultation revenue prior to the AI launch. Their participation therefore appears primarily driven by non-monetary motives such as altruism and recognition. Extrinsic-dominant counselors, in contrast, generated paid consultation revenue but did not contribute in the public forum before launch, indicating a stronger economic orientation toward monetized services. To sharpen identification, we exclude counselors active in both domains.

Table~\ref{tab:prosocial_mod} reports the subsample estimates. The responses diverge sharply. For intrinsic-dominant counselors, genAI integration significantly reduces participation intensity (Column~(1)), consistent with recognition dilution and intrinsic crowding-out. For extrinsic-dominant counselors, participation increases significantly after integration (Column~(2)), consistent with intensified competitive incentives in a more contested visibility environment.

Importantly, the decline among intrinsic-dominant counselors persists after controlling for patient posting volume (Columns~(3)–(4)). This finding indicates that these counselors reduce participation even when more opportunities arise, suggesting that their response reflects supply-side incentive changes rather than purely demand-driven dynamics. In contrast, the insignificant coefficient for extrinsic-dominant counselors after controlling for demand suggests that their increased effort scales proportionally with the expanded opportunity set rather than reflecting an absence of behavioral adjustment. Results remain robust when replacing the treatment indicator with the continuous AI activity measure (\textit{AIVolume}) in Columns~(5)–(6).

These findings provide direct evidence that intrinsic crowding-out and competitive intensification coexist following genAI integration. However, in a tiered professional platform, competitive incentives are structurally amplified because public participation affects paid-tier outcomes. As visibility becomes more contested, economically motivated counselors increase effort to defend or expand downstream market share. Intrinsic crowding-out thus emerges as a boundary condition affecting recognition-driven contributors, whereas competitive responses dominate among economically motivated professionals.

This heterogeneity distinguishes tiered professional platforms from single-tier voluntary communities, where AI entry may primarily depress participation \citep{RN1365}. In monetized ecosystems, genAI integration can simultaneously crowd out intrinsically motivated contributors while stimulating strategic effort among those exposed to competitive economic incentives.

\begin{table} 
\TABLE
{The Subsample Analysis\label{tab:prosocial_mod}
}
{\begin{tabular}{@{}lcccccc}
\hline \up
& (1) & (2)   & (3)&(4) & (5)&(6)\\
& ThoughtIntensity& ThoughtIntensity & ThoughtIntensity&ThoughtIntensity  & ThoughtIntensity&ThoughtIntensity  \\
 & (Intrinsic)&(Extrinsic) & (Intrinsic)&(Extrinsic)  & (Intrinsic)&(Extrinsic)  \\ \hline \up
\textit{Treat×PostLaunch}& -0.176***&                          0.029***
& -0.142***&-0.003
& &\\
                              & (0.041)&                          (0.003)
& (0.052)&(0.005)
& &\\
 \textit{AIVolume}& & & & & -0.025***&0.000
\\
 & & & & & (0.008)&(0.001)
\\
 \textit{TPostVolume}& & & -0.041&0.037***
& -0.025&0.035***
\\
 & & & (0.038)&(0.004)
& (0.040)&(0.004)
\\
Constant                      & 0.163***& 0.003**& 0.399*&-0.211***& 0.309&-0.199***
\\
                              & (0.019)& (0.001)& (0.218)&(0.023)& (0.229)&(0.024)
\\
Observations                  & 3,948& 57,058& 3,948&57,058& 3,948&57,058\\
R-squared                     & 0.156& 0.168& 0.157&0.170& 0.157&0.170\\
Year-Counselor FE& Yes                      & Yes                       & Yes                       &Yes                        & Yes                        &Yes                        \\
 Weekly FE& Yes                      &Yes                     & Yes                       &Yes                     & Yes                        &Yes                        \down \\ \hline
\end{tabular}}{Robust standard errors in parentheses. *** p\textless{}0.01, ** p\textless{}0.05, * p\textless{}0.1}
\end{table}

\subsection{Necessary Condition for the Demand-side Channel: Interference in the Non-AI \textit{Answers} Subforum}\label{sec:spillover_answers}

If the increase in counselor participation in the AI-enabled \textit{Thoughts} subforum is primarily driven by patient engagement expansion, this mechanism should generate cross-subforum spillovers. The two subforums partially overlap in functionality and serve similar informational needs, making them substitutable from patients’ perspective. By improving responsiveness and scalability, genAI increases the expected likelihood and speed of receiving replies in \textit{Thoughts}. When immediacy is valued, patients rationally shift attention toward the section with higher expected response quality and lower waiting costs.

Under this demand-reallocation mechanism, growth in patient activity in \textit{Thoughts} should be accompanied by a contraction of activity in the non-AI \textit{Answers} subforum. By contrast, if the baseline effects were driven by platform-wide shocks, such as seasonality or general growth in mental health awareness, posting activity would increase across both subforums.

We test this implication by re-estimating the posting-volume specification using patient posts in \textit{Answers} (\textit{APostVolume}) as the dependent variable. Column (1) of Table~\ref{tab:spill_answer} shows a statistically significant decline in \textit{Answers} posting following AI integration. This pattern supports engagement reallocation and mitigates concerns that our baseline findings reflect common shocks.

Demand reallocation also implies equilibrium supply adjustments. Counselors allocate limited time across subforums. When relative demand rises in \textit{Thoughts} but falls in \textit{Answers}, the opportunity cost of contributing to \textit{Answers} increases, creating incentives to shift effort toward the AI-enabled section. Consistent with this logic, Column (2) shows a decline in counselor participation intensity in \textit{Answers} after AI launch.

However, this decline may arise either from a direct technological effect of AI integration or from an indirect equilibrium response to shifting demand and competitive conditions across subforums. To distinguish between these explanations, we extend the specification by controlling for patient demand in both subforums and counselor supply in \textit{Thoughts}. This adjustment accounts for cross-forum substitution in effort allocation, as participation decisions depend on relative demand and visibility conditions across sections.

Once these cross-forum demand and supply factors are held constant, the treatment effect in \textit{Answers} becomes statistically insignificant (Column (3)). Similar results obtain when using the continuous AI activity measure (\textit{AIVolume}) in Columns (4)–(5). These findings indicate that the reduction in \textit{Answers} participation is not a direct effect of AI integration. Rather, it reflects an equilibrium reallocation of counselor effort in response to demand expansion and competitive asymmetry in the AI-enabled \textit{Thoughts} section. The cross-subforum spillovers therefore provide a necessary condition supporting the demand-side channel as a central driver of the main results.

\subsection{Necessary Condition for the Supply-side Channel: Spillovers to Paid Private Consultations}\label{sec:spillover_paid}

A necessary implication of the extrinsic competitive incentive mechanism concerns downstream economic outcomes in the paid tier. In OMHCs, participation in the public forum serves as a visibility and signaling channel through which counselors attract potential clients to paid consultations. GenAI integration reshapes this awareness landscape in two ways. First, AI-generated responses increase total content supply, diluting individual visibility if counselors do not adjust their behavior. Second, improved responsiveness reallocates patient attention toward the AI-enabled forum, intensifying competition for salience at the top of the conversion funnel.

If counselors remain inactive following AI integration, their relative exposure declines and they risk losing potential clients at the awareness stage. Even if cross-tier conversion rates remain unchanged, reduced visibility among new patients can translate into lower paid consultation demand. By contrast, counselors who strategically increase public participation may defend or expand their relative market share in the paid tier.

To examine this necessary condition, we conduct a heterogeneity analysis based on post-launch activity. Counselors are classified into inactive (no public contributions after integration) and active groups. Although not all active counselors increase participation relative to pre-AI levels, they collectively account for the aggregate posting increase documented earlier and therefore represent those who engage in competitive adjustment.

We focus on first-time patient transactions (\textit{FirstPatient}) as the outcome variable. New patients rely heavily on publicly observable information when selecting counselors, whereas repeat patients depend more on prior relationships and accumulated trust \citep{RN16,RN101,RN871}. First-time transactions thus provide a clean measure of how changes in public visibility translate into paid demand.

Applying the baseline specification to these subsamples yields sharply contrasting patterns. Inactive counselors experience a statistically significant decline in new patient demand following AI integration (Columns (1)–(2) in Table~\ref{tab:spill_demand1}). In contrast, active counselors exhibit an increase in new patient demand (Columns (3)–(4)).

These results provide direct economic evidence for the competitive incentive channel. In a tiered professional ecosystem, public participation operates as a strategic response to AI-induced visibility dilution. Counselors who fail to adjust lose downstream demand, whereas those who remain active are able to preserve or expand their market share. The spillover to the paid tier therefore confirms that AI integration intensifies competitive incentives rather than merely altering intrinsic motivation.

\begin{table} 
\TABLE
{The Spillover Effects of AI Launch in Non-AI Involved Section\label{tab:spill_answer}
}
{\begin{tabular}{@{}lccccc}
\hline \up
& (1) & (2)   & (3)&(4)&(5)\\
& APostVolume& AnswerIntensity& AnswerIntensity&AnswerIntensity&AnswerIntensity \\ \hline \up
\textit{Treat×PostLaunch}& -1.029***
&                          -0.011***& -0.003
&&\\
                              & (0.120)
&                          (0.004)& (0.005)
&&\\
 \textit{AIVolume}& & &  &-0.002***&-0.000
\\
 & & &  &(0.001)&(0.001)
\\
 \textit{APostVolume}& &  & 0.003&&0.003
\\
 & &  & (0.003)&&(0.003)
\\
 \textit{ThoughtIntensity}& &  & 0.859***
&&0.859***
\\
 & &  & (0.014)
&&(0.014)
\\
 \textit{TPostVolume}& & & -0.035***
& &-0.035***
\\
 & & & (0.005)
& &(0.005)
\\
Constant                      & 6.533***
& 0.069***& 0.243***
&0.069***&0.243***
\\
                              & (0.065)
& (0.002)& (0.043)
&(0.002)&(0.043)
\\
Observations                  & 94& 156,322& 156,322
&156,322
&156,322
\\
R-squared                     & 0.858& 0.275& 0.439
&0.275
&0.439
\\
Yearly FE& Yes                      & No& No&No&No \\
 Year-Counselor FE& No& Yes                      &  Yes                      &Yes                      &Yes                         \\
 Weekly FE& Yes                      &Yes                        & Yes                      &Yes                        &Yes                         \down  \\ \hline
\end{tabular}}{Robust standard errors in parentheses. *** p\textless{}0.01, ** p\textless{}0.05, * p\textless{}0.1}
\end{table}

\begin{table} 
\TABLE
{The Spillover Effects in Counselors' Fee-based Services\label{tab:spill_demand1}
}
{\begin{tabular}{@{}lcccc}
\hline \up
& (1) &(4)&(5)&(6)\\
& FirstPatient&FirstPatient&FirstPatient&FirstPatient\\
 & \multicolumn{2}{c}{Active Counselors}& \multicolumn{2}{c}{ Idle Counselors}\\ \hline \up
\textit{Treat×PostLaunch}& 0.072***&&-0.025***&\\
                              & (0.012)&&(0.005)&\\
 \textit{AIPostVolume}& & 0.009***& &-0.004***
\\
 & & (0.002)& &(0.001)
\\
Constant                      & 0.140***&0.147***&0.183***&0.153**
\\
                              & (0.005)&(0.005)&(0.002)&(0.062)
\\
Observations                  & 20,868&20,868&135,454&135,454
\\
R-squared                     & 0.479&0.479&0.474&0.474
\\
Year-Counselor FE& Yes&Yes&Yes&Yes\\
Weekly FE&                           Yes&Yes&Yes&Yes \down\\ \hline
\end{tabular}}{Robust standard errors in parentheses. *** p\textless{}0.01, ** p\textless{}0.05, * p\textless{}0.1}
\end{table}

\section{Conclusion}\label{sec:discussion}

Our findings clarify how genAI reshapes participation in tiered professional platforms. Rather than uniformly displacing human expertise, genAI alters the competitive and demand environment in which professionals allocate effort. When public participation is economically consequential—as in OMHCs where visibility affects paid consultation demand—AI-driven improvements in responsiveness can expand patient engagement and enlarge opportunities for human contribution. In this environment, counselors respond strategically: even as traditional social rewards become diluted, economically motivated professionals increase participation to maintain visibility and downstream demand. The result is an equilibrium in which automated scalability and sustained human participation coexist. This coexistence is central to platform sustainability: genAI enhances access and responsiveness, while human counselors remain essential for addressing complex emotional needs and generating the paid services that finance the ecosystem. More broadly, our results suggest that the impact of genAI on professional communities depends critically on institutional design, highlighting a governance logic in which AI complements human contributors by reshaping incentives rather than replacing expertise.

Theoretically, our study contributes to three streams of research. First, we extend the OMHC literature, which has predominantly focused on patient-side behaviors and health outcomes, by systematically analyzing how platform technologies reshape expert supply. We show that internal AI integration does not simply substitute for professional contributions; rather, it reconfigures participation incentives in a tiered ecosystem where visibility affects economic returns. By uncovering heterogeneous responses across intrinsically and extrinsically motivated counselors, we illuminate the strategic dimension of professional participation that has received limited attention.

Second, we contribute to research on managing online content platforms by moving beyond traditional incentive levers. Prior work emphasizes feedback systems, reputation devices, or monetary rewards as tools for motivating contributors. In contrast, genAI directly produces content, thereby altering both the competitive environment and demand conditions. Our results demonstrate that AI can simultaneously dilute intrinsic recognition while intensifying market-based competition and expanding demand. This dual effect highlights that technological interventions reshape equilibrium behavior not only through cost changes but also through visibility-based market dynamics.

Third, we enrich the growing literature on genAI in online communities by shifting the focus from substitution toward structured complementarity. Much of the existing evidence documents negative effects of external AI tools on voluntary participation \citep{RN1363,RN1270}. By examining an internally deployed AI agent in a tiered professional setting, we show that AI can expand user engagement and enlarge the opportunity set for differentiated human expertise. In such environments, increased engagement interacts with competitive incentives tied to downstream economic outcomes, allowing demand expansion and strategic effort to outweigh intrinsic crowding-out. This finding identifies an important boundary condition for prior substitution-based results and highlights the role of institutional structure in shaping human–AI coexistence.

Our study also offers important managerial implications for platform operators overseeing hybrid ecosystems that integrate both free and paid services. First, AI deployment should be evaluated from a two-sided perspective. Improving responsiveness through internal AI can stimulate user engagement and indirectly sustain professional participation, rather than mechanically displacing experts. When AI expands demand and increases interaction volume, it can complement human contributors by enlarging their effective audience.

Second, because AI integration simultaneously dilutes recognition and intensifies competition, platforms must carefully manage visibility mechanisms. Enhanced attribution systems, differentiated reputation signals, or algorithmic highlighting of high-quality human contributions may mitigate intrinsic crowding-out while preserving productive competitive incentives. 

Third, the economic linkage between public visibility and paid conversion is central to sustaining participation in tiered settings. Platforms can strengthen this linkage through transparent recommendation systems, analytics dashboards, and targeted matching tools that make the returns to public engagement more salient. By reinforcing the connection between free-tier effort and paid-tier demand, platforms can channel competitive pressure into productive participation rather than disengagement.

Finally, AI-induced participation dynamics are unlikely to be homogeneous across contributors. Responses may vary depending on underlying motivations and exposure to economic incentives. Platforms should therefore adopt differentiated and adaptive governance mechanisms to accommodate these varied responses and ensure that AI integration strengthens, rather than undermines, long-term platform vitality.

Several limitations suggest avenues for future research. First, our analysis relies on observable public activity and cannot capture unobserved behaviors such as private communications, offline consultations, or counselors’ use of external AI tools. Future work could employ mixed-method approaches or platform collaborations to examine these dimensions. Second, our setting is situated in the mental health domain, where relational and emotional concerns may moderate substitution effects \citep{RN577,RN1034,RN1154,RN452}. The coexistence of AI and human expertise may differ in domains where informational accuracy dominates relational trust. Exploring how institutional structure conditions human–AI complementarity remains a promising direction for future research.

\bibliographystyle{informs2014} 
\bibliography{references} 





  


\newpage
\begin{APPENDICES}
\section{The PSM Results for Cross-platforms counselors} \label{sec:psm_appendix}
\setcounter{table}{0}
\setcounter{figure}{0}
\renewcommand\thefigure{\Alph{section}\arabic{figure}}    
\renewcommand\thetable{\Alph{section}\arabic{table}}
Figure \ref{fig:cross_psm} shows the distribution of propensity score for both treatment and control group before and after matching.
 \begin{figure}[!hb]
     \FIGURE
     {\includegraphics[width=\textwidth]{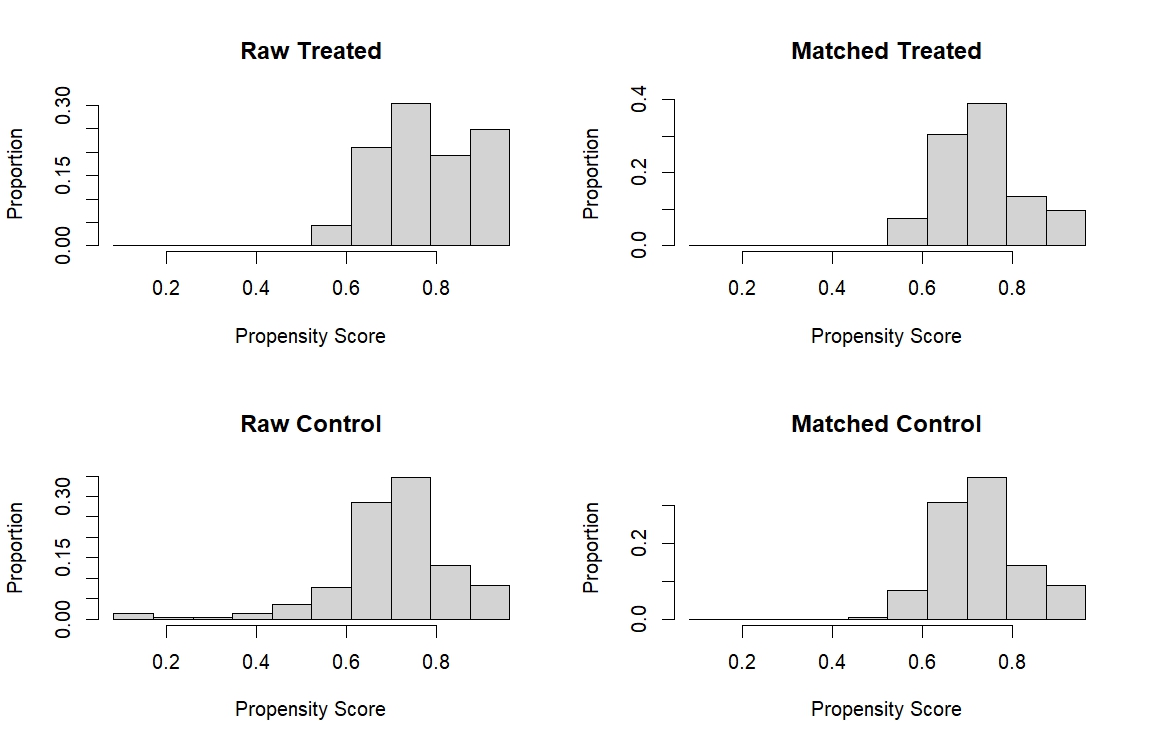}} 
{Distribution of Propensity Score Before and Matching. \label{fig:cross_psm}}
{}
\end{figure}
\end{APPENDICES}

%
%
%







\end{document}